\documentclass{aastex631}

\usepackage{bm}
\usepackage{xcolor}
\usepackage{multibib}
\newcites{main}{References}
\newcites{appendix}{References (Appendix)}
\usepackage{subfigure}
\usepackage{soul}
\usepackage{tcolorbox}
\usepackage{amsmath}
\usepackage{verbatim}
\newcommand{\beq}{\begin{equation}}
\newcommand{\eeq}{\end{equation}}
\newcommand{\bseq}{\begin{subequations}}
\newcommand{\eseq}{\end{subequations}}
\newcommand{\bary}{\begin{eqnarray}}
\newcommand{\eary}{\end{eqnarray}}
\newcommand{\bwt}{\begin{widetext}}
\newcommand{\ewt}{\end{widetext}}

\begin{document}
\title{Very High-energy Afterglow Emission of GRB 190829A: Evidence for Its Hadronic Origin?}

\author{Sarira Sahu}
\email{sarira@nucleares.unam.mx}
\affiliation{Instituto de Ciencias Nucleares, Universidad Nacional Aut\'onoma de M\'exico, \\
Circuito Exterior, C.U., A. Postal 70-543, 04510 Mexico DF, Mexico}

\author{Isabel Abigail Valadez Polanco}
\email{abivaladez@gmail.com}
\affiliation{Facultad de Ingeniería, Universidad Aut\'onoma de Yucat\'an, \\
Industrias No Contaminantes S/N, Sin Nombre de Col 27, M\'erida, Yucat\'an, M\'exico}

\author{Subhash Rajpoot}
\email{Subhash.Rajpoot@csulb.edu}
\affiliation{Department of Physics and Astronomy, California State University,\\ 
1250 Bellflower Boulevard, Long Beach, CA 90840, USA}

\begin{abstract}
The detection of multi-TeV gamma-rays from the afterglow phase of GRB 190829A by High Energy Stereoscopic System (H.E.S.S.) telescope is an addition to the already existing list of two more GRBs observed in the very high energy (VHE) gamma-rays in recent years. Jets of blazars and GRBs have many similarities and the photohadronic model is very successful in explaining the VHE gamma-ray spectra from the high energy blazars. Recently, the photohadronic model has been successfully applied to study the sub-TeV gamma-rays from the afterglow phases of GRB 180720B and GRB 190114C. We employed this model again to explain the VHE spectra observed for the two consecutive nights from GRB 190829A. We show that the spectra of GRB 190829A can be due to the interactions of high energy protons with the synchrotron self-Compton photons in the forward shock region of the GRB jet, similar to the low emission state of the VHE flaring events of high energy blazars. We speculate that, if in future, it is possible to observe the VHE gamma-ray spectra from  nearby GRBs in their afterglow phases, then some of them could only be explained by employing two different spectral indices. If confirmed, such VHE spectra could be interpreted as a result of the interactions of the high energy protons with the photons, both from the synchrotron background and the synchrotron self-Compton background in the forward shock region.
\end{abstract}


\section{Introduction} \label{sec:intro}

The birth of Gamma-ray  bursts (GRBs) signal the formation of stellar-mass black holes or fast rotating magnetars either from the collapse of massive stars or from the merger of very compact binaries (\cite{Piran:2004ba,Kumar:2015pr} and Refs. therein).  The afterglow in GRBs takes place soon after the prompt phase and the standard model to understand them is the external shock model. According to this model, the relativistic ejecta (also called fireball) gets decelerated by the circumstellar medium and as a consequence two shocks are developed. The one which lasts longer is the forward shock that is still penetrating through the medium while the short-lived second one is the reverse shock propagating into the fireball. Gamma-rays above $\sim 100$ {\it MeV} are believed to be produced from electron synchrotron in the afterglow shocks \citep{Sari:2000zp,2009MNRAS.400L..75K,2010MNRAS.403..926G,2010ApJ...712.1232W}. Also, joint observations of X-rays by {\it Swift}-X-ray Telescope (XRT) and gamma-rays by {\it Fermi} Large Area Telescope (LAT) from many GRBs are consistent with a single spectral component \citep{Ajello:2018}.
However, it is difficult to produce photons above a few {\it GeV} in the context of synchrotron mechanism unless an unrealistically large bulk Lorentz factor is employed \citep{Razzaque:2009rt}. 
The fireball model also predicts the origin of GeV-TeV photons and the emission can last from minutes to several hours \citep{Piran:2004ba,Kumar:2015pr}. The synchrotron self-Compton (SSC) process is the favored one and widely used mechanism to interpret the very high energy (VHE $> 100$ {\it GeV}) photons in the afterglow era, where the relativistic electrons upscatter their own synchrotron radiation\citep{1997ApJ...485L...5W,2001ApJ...559..110Z,Sari:2000zp,Kumar:2015pr}. In this scenario, the SSC emission takes place from a constant density circumburst medium or from the Comptonization of X-ray photons in the afterglow shock \citep{Wang:2019zbs,Derishev:2019cgi}.

Recently, the VHE photons were detected from the afterglow phases of the  GRB 180720B by Major Atmospheric Gamma Imaging Cherenkov (M.A.G.I.C.) telescopes and from the GRB 190114C and GRB 190829A by H.E.S.S. telescope \citep{Arakawa:2019cfc,Acciari:2019dxz,HESS:2021dbz}. The detection of VHE photons from GRBs by the ground based Imaging Air Cherenkov Telescopes (IACTs), like the M.A.G.I.C. and the H.E.S.S., is a new development in the study of GRBs in the VHE $\gamma$-ray regime and will provide vital information on the physical processes responsible for the radiation mechanisms and particle acceleration in an extreme environment \citep{2006RPPh...69.2259M,Kumar:2015pr, Fraija:2019rag}.

GRB 190829A was first detected by the Gamma-Ray Burst Monitor (GBM) on board the {\it {Fermi}} gamma-ray Space Telescope on 29 August 2019 at 19:55:53 universal time (UT, $T_0$) \citep{GCN:25575}. Subsequently, it was observed and followed in the multiwavelength by several other telescopes \citep{HESS:2021dbz}. The afterglow was also observed by ground based optical, infra-red and radio telescopes starting 1318 s after the GBM trigger and observations were continued for several days afterwards. The measured redshift of the host galaxy is $z=0.0785\pm 0.0005$ \citep{AstroTel:13052,GCN:25565}, making this the nearest GRB detected in VHE so far. The H.E.S.S. telescope observed multi-TeV gamma-rays from 4.3 to 55.9 hours after the prompt emission with a direction consistent with the location of the source. GRB 190829A has some observational peculiarities when compared with the other two GRBs, GRB 180720B and GRB 190114C, that have been recently observed in VHE. It has two episodes of prompt emission, separated by a quiescent time gap of about 40 s, and both of them are of different nature. 
Also the isotropic equivalent luminosity of GRB 190829A is $L_{iso}\sim 10^{49}\, \mathrm{erg\, s^{-1}}$ which is smaller than the typical long GRBs, and possibly belongs to low-luminosity GRB (LLGRB) category \citep{Chand:2020wqt}. Association of the GRB 190829A with the broad-line type-Ic supernova SN 20190yw has been established \citep{Hu:2020xhu}.

Observations of VHE photons at late times ($>100$s) are difficult to comprehend due to the substantial decrease in the bulk Lorentz factor of the shock wave. Thus, various radiation mechanisms mostly based on the Inverse Compton scattering, Proton synchrotron model and their variants are proposed to explain the origin of VHE $\gamma$-rays during the afterglow phase.
The SSC \citep{Derishev:2019cgi,Wang:2019zbs,Vurm:2016qqi} and/or the external inverse-Compton (EIC) \citep{2020arXiv201207796Z} scenarios are used to explain the sub-TeV emission from GRBs.
The extremely high energy protons in the magnetic field of the afterglow shock can produce the observed gamma-ray spectrum through the synchrotron process\citep{1998ApJ...502L..13T,Razzaque:2009rt}. Also, the interaction of these energetic protons with the background photons in the afterglow shock  can produce the observed VHE gamma-rays through the photopion process \citep{Asano:2012jr,Sahu:2020dsg}.  The advantages and disadvantages of leptonic and hadronic models are reviewed in \citep{Kumar:2015pr}.

Several studies on the emission mechanisms in blazars and GRBs  show many similarities \citep{Nemmen:2012rd,Wang:2010nr,Wu:2015opa}. 
 It is observed that the GRB afterglows have the same radiation mechanism as the BL Lac objects \citep{Wang:2010nr}. A similar correlation of the synchrotron luminosity and Doppler factor between GRBs and AGNs has been found \citep{Wu:2011xrt}. Also, the relativistic jets in AGNs and GRBs have
a similar energy dissipation efficiency \citep{Nemmen:2012rd}.
All the above studies show that, despite order of magnitude differences in their masses and bulk Lorentz factors, the jets in GRBs and blazars have many characteristics in common. Thus, it is imperative to study the VHE emission mechanisms in the afterglow phases of GRBs by using the common mechanisms and processes that are being used to study the multi-TeV flaring of high energy blazars.

In studying the multi-TeV flaring from high-energy peaked blazars (HBLs) we have made use of the photohadroinc model \citep{Sahu:2019scf,Sahu:2019lwj,Sahu:2019kfd}. In this model, within the blazar jet, the Fermi accelerated high energy protons interact with the background seed photons through the process $p\gamma\rightarrow \Delta^+$ and the subsequent decay of the $\Delta$-resonance produces VHE $\gamma$-rays through intermediate neutral pion decays. This model is very successful in explaining the VHE $\gamma$-ray spectrum from several HBLs. Recently, keeping in mind the similarity between the emission mechanisms in GRBs and blazars, we have applied the photohadronic model to explain the sub-TeV emission from the GRB 190114C and GRB 180720B and explain the VHE $\gamma$-ray spectra of both these GRBs extremely well \citep{Sahu:2020dsg}. We have also shown that the VHE spectrum of GRBs 190114C is due to the interaction of Fermi accelerated high energy protons with the SSC photons while the spectrum of GRB 180720B is from the interaction of high energy protons with the synchrotron photons in the external forward shock region. 

In this work we wish to exploit the success of the photohadronic model once again to explain the VHE afterglow emission from the GRB 190829A.

\section{Photohadronic scenario}
The kinematical condition to produce VHE photons from the $p\gamma$ interaction is given by \citep{Sahu:2019lwj}
\begin{linenomath*}
\beq
E_{\gamma} \epsilon_{\gamma} = \frac{0.032\, \Gamma\,{\cal D}}{(1+z)^{2}} \, \mathrm{{\it GeV}^2},
\label{KinemCond}
\eeq
\end{linenomath*}
where the observed VHE photon energy is $E_{\gamma}$ and $\epsilon_{\gamma}$ is the background seed photon energy.
$\Gamma$ and ${\cal D}$ are the bulk Lorentz factor and the Doppler factor respectively. The observed HBLs and GRBs have their jets beaming towards the observer on earth, so we have $\Gamma\simeq{\cal D}$. $z$ is the redshift of the source. In the photohadronic process, the observed VHE photon carries about 10\% of the proton energy, i.e., $E_{\gamma}\simeq 0.1\, E_p$, where $E_p$ is the proton energy.

The VHE photons from extragalactic sources get attenuated due to their interactions with the extragalactic background light (EBL) by producing $e^+e^-$ pairs \citep{1992ApJ...390L..49S,2012Sci...338.1190A}. This reduces the VHE flux by a factor $e^{-\tau_{\gamma\gamma}}$, where $\tau_{\gamma\gamma}$ is the optical depth for the lepton pair production process which depends on $E_{\gamma}$ and $z$. In the photohadronic model, the observed VHE $\gamma$-ray flux $F_{\gamma}$ is proportional to $E^2_{\gamma}dN/dE_{\gamma}$ and is given by (\cite{Sahu:2019lwj} and Refs. therein)
\begin{linenomath*}
\beq
F_{\gamma}(E_{\gamma}) = F_0
\left (   \frac{E_{\gamma}}{{\it TeV}}
\right )^{-\delta+3} e^{-\tau_{\gamma\gamma}}
= F_{\gamma,int}\, e^{-\tau_{\gamma\gamma}} .
\label{FluxObs}
\eeq
\end{linenomath*}
The normalization factor $F_0$ can be fixed from the observed spectrum. The spectral index $\delta=\alpha+\beta$ is the free parameter in the model and $F_{\gamma,int}$ is the intrinsic flux.
The Fermi accelerated protons in the jet have a power-law profile $dN/dE_p\propto E^{-\alpha}_p$ and the spectral index $\alpha\ge 2$. Here we use the generally accepted value of $\alpha=2.0$ \citep{Dermer:1993cz}. $\beta$ is the spectral index of the background seed photons and it is further observed that for HBLs the seed photon flux is also a power-law $\Phi_{\gamma}\propto  \epsilon^{\beta}_{\gamma}\propto E^{-\beta}_{\gamma}$. However, for GRBs, the $\beta$ value can be either positive or negative as shown in \citep{Sahu:2020dsg}. Also, the sign of $\beta$ locates the seed photon background in the synchrotron regime ($\beta < 0$) or in the SSC regime ($\beta > 0$). By fitting the observed VHE spectrum, the value of $\delta$ is fixed and it automatically fixes the value of $\beta$. For HBLs, we have shown that the value of $\delta$ is always in the range $2.5\le \delta \le 3.0$ \citep{Sahu:2019kfd}. 

\begin{figure}[ht!]
\plotone{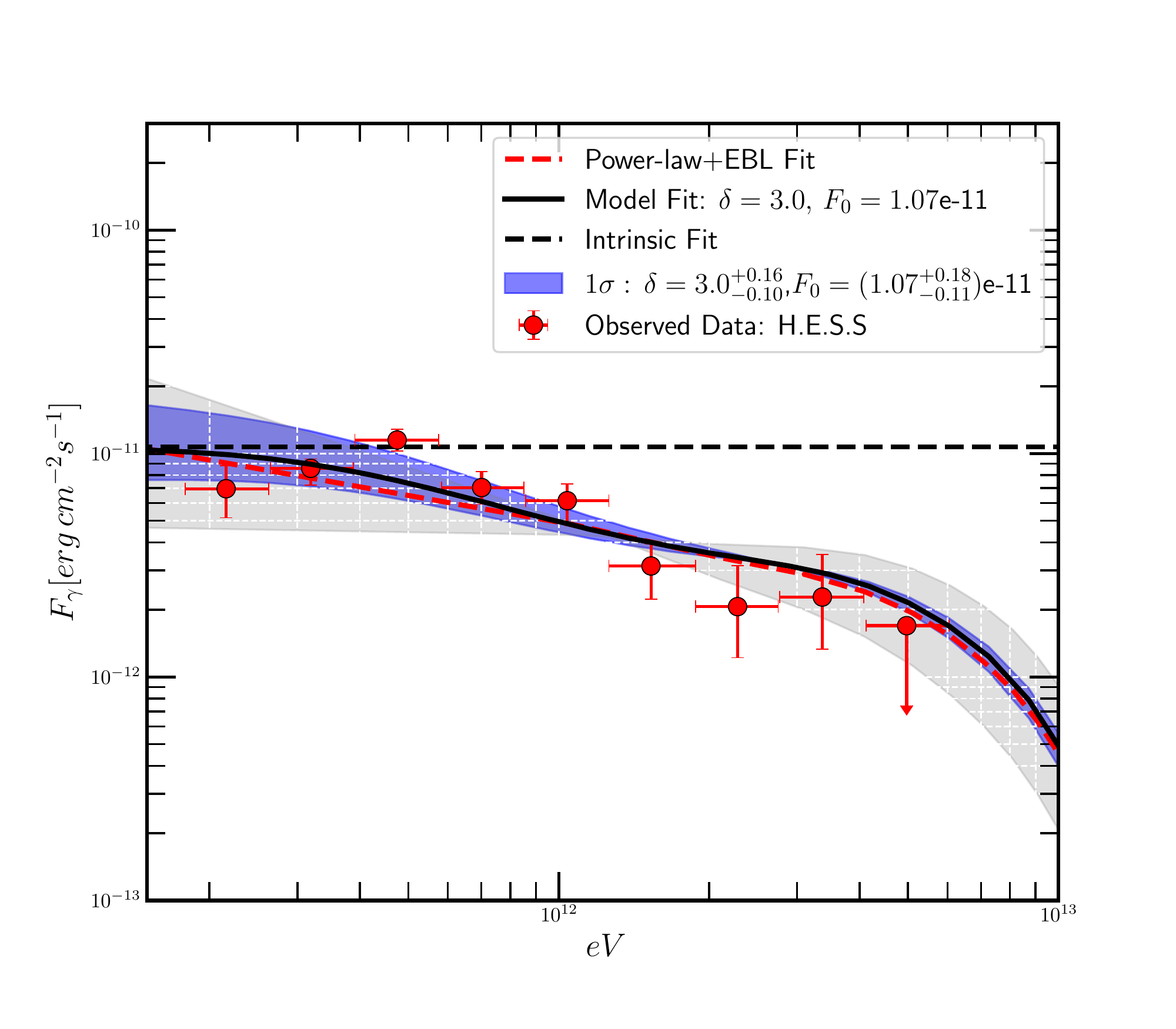}
\caption{
  The multi-TeV $\gamma$-ray spectrum observed by H.E.S.S. telescope from GRB 190829A on the first night is fitted with the photohadronic model. The normalization constant $F_0$ is expressed in units of $\mathrm{erg\,cm^{-2}\,s^{-1}}$. The blue shaded region corresponds to
  $F_0=(1.07^{+0.18}_{-0.11})\times 10^{-11}\, \mathrm {erg\, cm^{-2}\, s^{-1}}$ and $\delta=3.0^{+0.16}_{-0.10}$. The intrinsic flux for the photohadronic model is also shown.  The photohadoronic fit is 
compared with the power-law+EBL fit, where $\gamma^{int}_{VHE}=2.06$, $E_0=0.556$ {\it TeV} and $F_{PW}=1.12\times 10^{-11}\, \mathrm{erg\, cm^{-2}\, s^{-1}}$ are used. The light grey shaded region is between $\gamma^{int}_{VHE}=2.42$ (upper edge) and $\gamma^{int}_{VHE}=1.7$ (lower edge) where the statistical and the systematic errors are taken into account.
\citep{HESS:2021dbz}. 
\label{fig:figure1}}
\end{figure}
\begin{figure}[ht!]
\plotone{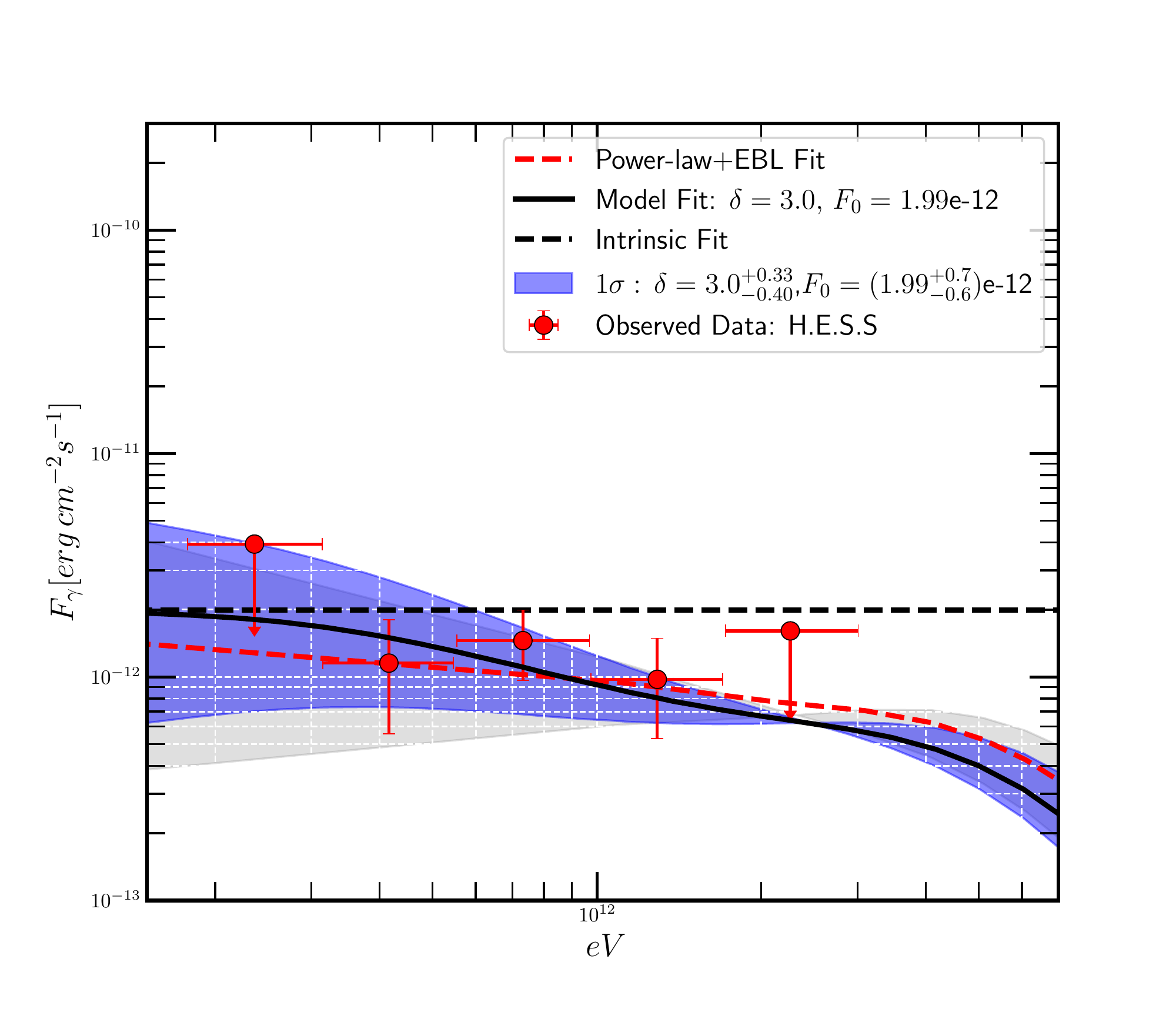}
\caption{
  The VHE $\gamma$-ray spectrum observed by H.E.S.S. telescope from GRB 190829A (second night) is fitted with the photohadronic model and compared with the power-law+EBL fit ($\gamma^{int}_{VHE}=1.86$, $E_0=0.741$ {\it TeV} and $F_{PW}=2.03\times 10^{-12}\, \mathrm{erg\, cm^{-2}\, s^{-1}}$ are used ). The light grey shaded region is between $\gamma^{int}_{VHE}=2.29$ (upper edge) and $\gamma^{int}_{VHE}=1.43$ (lower edge) when the statistical and the systematic errors are taken into account  \citep{HESS:2021dbz}. The intrinsic flux for the photohadronic model is a flat curve. The normalization constant $F_0$ is expressed in units of $\mathrm {erg\,cm^{-2}\,s^{-1}}$. The blue shaded region corresponds to $F_0=(1.99^{+0.70}_{-0.60})\times 10^{-12}\,
\mathrm {erg\, cm^{-2}\, s^{-1}}$ and $\delta=3.0^{+0.33}_{-0.40}$.}
\label{fig:figure2}
\end{figure}

\section{results}
The H.E.S.S. collaboration observed the afterglow of the GRB 190829A for three consecutive nights \citep{HESS:2021dbz}. In the first night it was observed, starting at $T_0+4.3$ hours, for 3.6 hours. In the second night the observation started at $T_0+27.2$ hours and continued for 4.7 hours. In the third night the  beginning time was at $T_0+51.2$ hours and lasted for 4.7 hours. Although H.E.S.S. telescope observed gamma-rays in all the nights, the signal in the last night was too weak to construct the spectrum. In the first night the observed spectrum was in the energy range $0.18\, {\it TeV} \le E_{\gamma} \le 3.3\, {\it TeV}$ and in the second night the spectrum was in the energy range $0.18\, {\it TeV} \le E_{\gamma} \le 1.4\, {\it TeV}$. The differential photon spectra of both the nights are perfectly compatible with a power-law with the EBL correction. However, it is shown that the values of the spectral indices are not the same for both the nights \citep{HESS:2021dbz}.

We fitted the spectra of both the nights using the photohadronic model. It is observed that the spectrum of the first night can be fitted with the spectral index $\delta=3.0$ and the normalization factor $F_0=1.07\times 10^{-11}\, \mathrm{erg\, cm^{-2}\, s^{-1}}$ as shown in Fig. \ref{fig:figure1}. The statistical significance of this fit is $91.7\%$. Here, we have used the $\chi^2$ fit to calculate the statistical significance. To account for the EBL effect, we have used the EBL model of \cite{Franceschini:2008tp}. However, the EBL model of \cite{Dominguez:2010bv} also gives similar result. Our fit shows that the decrease in the flux is due to the EBL correction only. For $E_{\gamma}=3.3$ {\it TeV} the flux is depleted by about 73\%. 

In the photohadronic model $F_0$ and $\delta$ serve as two free parameters. Although a best fit is obtained for $\delta=3.0$ and $F_0=1.07\times 10^{-11}\, \mathrm{erg\, cm^{-2}\, s^{-1}}$, it is important to estimate the errors in both these parameters. 
By keeping one parameter free while the other one is frozen at its optimum value
we estimated the $1 \sigma$ error on these parameters which give $F_0=(1.07\pm 0.23)\times 10^{-11}\, \mathrm{erg\, cm^{-2}\, s^{-1}}$ and $\delta=3.0 \pm 0.2$. 
Also, with the simultaneous variation of the two free parameters, the $1 \sigma$ confidence interval defined by $\chi^2_{min}+2.3$ \citep{Zyla:2020zbs}, is calculated and the errors in the parameters are  respectively
$F_0=(1.07^{+0.18}_{-0.11})\times 10^{-11}\, \mathrm {erg\, cm^{-2}\, s^{-1}}$ and $\delta=3.0^{+0.16}_{-0.10}$. The blue shaded region in Fig. \ref{fig:figure1} corresponds to these values of $F_0$ and $\delta$.

We have also compared our fit with the power-law plus EBL fit. For the power-law plus EBL fit the flux is defined by $F_{\gamma}=F_{PW} (E_{\gamma}/E_0)^{-\gamma^{int}_{VHE}+2}\,\mathrm {e^{-\tau_{\gamma\gamma}}}$ and the central values of $\gamma^{int}_{VHE}=2.06$, $E_0=0.556$ {\it TeV} and $F_{PW}=1.12\times 10^{-11}\, \mathrm{erg\, cm^{-2}\, s^{-1}}$ are used \citep{HESS:2021dbz} and the statistical significance of this fit is $86.3\%$. Taking into account the statistical and the systematic errors, we have shown the light grey region for this fit (with $\gamma^{int}_{VHE}=2.42$, the upper edge and $\gamma^{int}_{VHE}=1.7$, the lower edge of the grey region).
According to the classification of the VHE flaring events of HBLs \citep{Sahu:2019kfd}, $\delta=3.0$ corresponds to low emission state and the intrinsic flux is a flat curve independent of energy. This value of $\delta$ corresponds to the seed photon spectral index $\beta=1.0$. In the context of HBL flaring it is shown that positive value of $\beta$ corresponds to seed photons in the SSC regime. Thus, to produce this VHE spectrum, the high energy protons in the energy range $1.8\, TeV \le E_p \le 33.0\, {\it TeV}$ should interact with the SSC photons in the forward shock region.

In the photohadronic model, the optical depth for the $\Delta^+$ production during the afterglow is $\tau_{p\gamma}=n^{\prime}_{\gamma}\sigma_{\Delta}R^{\prime}$, where $n^{\prime}_{\gamma}$ is the comoving SSC photon density in the forward shock region and $R^{\prime}\sim 2.9\times 10^{16}$ cm is the comoving distance from the central engine \citep{HESS:2021dbz}. For the above process we assume a mild efficiency by taking  $\tau_{p\gamma} \,< 1$ and this gives $n^{\prime}_{\gamma} < 7\times 10^{10}\,\mathrm{cm^{-3}}$. Also, the $e\gamma$ interaction is taking place in the same background and we determine $n^{\prime}_{\gamma} < 5.2\times 10^{7}\,\mathrm{cm^{-3}}$. 
So, by taking $n^{\prime}_{\gamma} \sim 10^{8}\,\mathrm{cm^{-3}}$ we get $\tau_{p\gamma} \,\sim 1.5\times 10^{-3}$. The integrated VHE photon flux in the energy range $0.18\, TeV \le E_{\gamma} \le 3.5\, TeV$ is $F_{\gamma}\simeq 1.8\times 10^{-11} \mathrm{erg\, cm^{-2}\, s^{-1}}$ and corresponds to luminosity
$L_{\gamma} \simeq 2.6\times 10^{44}\,\mathrm{erg\, s^{-1}}$. The isotropic-equivalent energy emitted in VHE during the 3.6 hours period in the first night is determined as $E^{iso}_{\gamma}\simeq 3.4\times 10^{48}\, \mathrm{erg}$. By taking $\tau_{p\gamma} \,\sim 1.5\times 10^{-3}$, the proton luminosity is estimated to be 
$L_p\simeq 1.3\times 10^{48}\,\mathrm{erg\, s^{-1}}$. Thus we conclude that a typical LLGRB has much lower energy output than a long GRB \citep{Acciari:2019dxz,Acciari:2019dbx,Sahu:2020dsg}.

On the second night the GRB jet is decelerated compared to the first night and the VHE photon energy is also decreased. So, if the spectrum of the first night is fitted with $\delta=3.0$ corresponding to a low emission state, the spectrum of the second night can not be fitted with $\delta < 3.0$ as this corresponds to high or very high emission state depending on the value of $\delta$ \citep{Sahu:2019kfd}. However, the VHE spectrum of the second night is also fitted with the photohadronic model and a good fit is obtained for $\delta=3.0$ and  $F_0=1.99\times 10^{-12}\, \mathrm{erg\, cm^{-2}\, s^{-1}}$. As there are only three points, we do not calculate the statistical significance of it. Again, as $\delta=3.0$, this is also in the low emission state and the background photon flux is proportional to $\epsilon_{\gamma}$. The intrinsic spectrum is a flat curve and is independent of energy. We also estimate the $1 \sigma$ errors in the parameters $\delta$ and  $F_0$ as before for the VHE spectrum of the second night. As this spectrum has only three data points the estimated $1 \sigma$ errors are large and we obtain $F_0=(1.99\pm 0.74)\times 10^{-12}\, \mathrm{erg\, cm^{-2}\, s^{-1}}$ and $\delta=3.0\pm 0.35$ respectively by varying one parameter while the other one is frozen at its optimum value. With simultaneous variation of the free parameters $F_0$ and $\delta$, the $1\, \sigma$ error in these parameters are given by $F_0=(1.99^{+0.70}_{-0.60})\times 10^{-12}\,
\mathrm {erg\, cm^{-2}\, s^{-1}}$ and $\delta=3.0^{+0.33}_{-0.40}$ respectively.
In Fig. \ref{fig:figure2}, the blue shaded region corresponds to these errors in $F_0$ and $\delta$ respectively. Also, the photohadronic fit is compared with the the power-law plus EBL fit. For $E_{\gamma} < 1\, {\it TeV}$, the photohadronic fit is different from the power-law plus EBL fit.


In the photohadronic scenario, the VHE spectrum is produced from the interactions of high energy protons in the energy range $1.8\, TeV \le E_p \le 14\, TeV$ with the SSC seed photons in the forward shock region as in the first night. Although our photohadronic fit (for $\delta=3.0$) is similar to the power-law plus EBL fit (here, the central values of $\gamma^{int}_{VHE}=1.86$, $E_0=0.741$ TeV and $F_{PW}=2.03\times 10^{-12}\, \mathrm{erg\, cm^{-2}\, s^{-1}}$ are used), our spectrum falls slightly faster than the latter fit. Also, below 1 TeV, the behavior of both the fits are slightly different as can be seen from Fig. \ref{fig:figure2}. 
The integrated VHE flux in the energy range $0.18\, TeV \le E_{\gamma} \le 1.4\, TeV$ is $F_{\gamma}=3.0\times 10^{-12}\, \mathrm{erg\, cm^{-2}\, s^{-1}}$ and the luminosity is 
$L_{\gamma} \simeq 4.4\times 10^{43}\,\mathrm{erg\, s^{-1}}$. In the second night the VHE emission was observed for 4.7 hours. The isotropic-equivalent energy emitted during this period is $E^{iso}_{\gamma}\simeq 7.5\times 10^{47}\, \mathrm{erg}$. Taking $R^{\prime}\sim 10^{17}$ cm and assuming a mild efficiency for the $p\gamma$ process, we get $n^{\prime}_{\gamma} < 2\times 10^{10}\,\mathrm{cm^{-3}}$. By taking the SSC photon density $n^{\prime}_{\gamma} \sim 10^{7}\,\mathrm{cm^{-3}}$, we get $\tau_{p\gamma} \,\sim 5\times 10^{-4}$ and the corresponding proton luminosity is $L_p\simeq 6.6\times 10^{47}\,\mathrm{erg\, s^{-1}}$. Thus, in the second night $L_p$ and $E^{iso}_{\gamma}$ are about 51\% and 22\% of the first night values respectively.

From the multiwavelength modeling of the X-ray to VHE $\gamma$-ray data of GRB 190829A, the value of $\Gamma$ is found to be $\Gamma=4.7$ for the first night and $\Gamma=2.6$ for the second night respectively \citep{HESS:2021dbz}. Using these values of $\Gamma$ in the photohadronic model, we estimate the seed photon energies for both the nights. By taking $E_{\gamma}= 3.3$ TeV as the highest photon energy in the first night, the seed photon energy is found to be $\epsilon_{\gamma}=184$ keV. Similarly, the highest photon energy for the second night  is $E_{\gamma}=1.4$ TeV, and this gives $\epsilon_{\gamma}=56$ keV. The calculated values of $\epsilon_{\gamma}$ for both these nights are low and belong to the synchrotron regime \citep{Sahu:2019kfd}. According to the photohadronic scenario, since the seed photon energy $\epsilon_{\gamma}$ should be in the SSC regime, we expect that the $\epsilon_{\gamma}$ value should be in the $\mathcal {O}$ (few {\it MeV}) range. Thus, by taking $\epsilon_{\gamma}\sim 1$ {\it MeV} for the first night, we get $\Gamma\sim 11$, which is a mild value and compatible with the seed photon energy in the SSC regime. Similarly we can also estimate $\Gamma$ for the second night. However, it turns out that the VHE afterglow of the long GRBs, GRB 190114C and GRB 180720B have large $\Gamma$ \citep{Sahu:2020dsg} as their synchrotron spectra get extended to $\sim 100$ {\it MeV}. 

The photohadronic model works well for $E_{\gamma}\gtrsim 100$ {\it GeV} and to produce gamma-rays below this energy the required seed photon energy is $\epsilon_{\gamma} > 72\, {\it MeV}$ which is in SSC regime. The density of these seed photons is small which makes the $\Delta$ production inefficient from $p\gamma$ interactions. Also for $E_{\gamma}\,<\,100 \, {\it GeV}$ the leptonic processes are much more important than the photohadronic process and the latter can be neglected to explain the spectral energy distribution in this energy regime. Also, in the photohadronic scenario, for $E_{\gamma}\gtrsim 100$ {\it GeV}, the main contribution to the VHE spectrum comes from the $p\gamma$ process. Hence, leptonic contributions to the VHE spectrum are neglected.

\section{Conclusions}

GRB 190829A is the third GRB observed in the VHE $\gamma$-rays and the second GRB observed by the H.E.S.S. telescope in the afterglow epoch between $T_0+4.3$ to $T_0+56$ hours. Due to many similarities between the HBL flaring and the GRB emission, we employed the successful photohadronic model to study the VHE afterglow of the GRB 190829A. The observed multi-TeV spectra of two consecutive nights from GRB 190829A can be explained well by the photohadronic model with the inclusion of the EBL correction. However, the $\gamma\gamma$ absorption within the jet is negligible in our model.  We have shown that the interactions of high energy protons with the SSC seed photons in the forward shock region seems to be responsible for the production of VHE $\gamma$-rays observed in both the nights. These spectra are similar to the spectrum observed from HBLs during low state emission corresponding to a spectral index $\delta=3.0$. Also this value of $\delta$ implies that the SSC photon flux depends linearly on $\epsilon_{\gamma}$. The intrinsic spectra are flat and independent of $E_{\gamma}$ for both nights.
We have also shown that the bulk Lorentz factor of the GRB should be $\Gamma\sim 11$ to explain the VHE spectra. Finally, it is important to mention that the photohadronic model in its simplest form as used previously,
could explain very well the sub-TeV afterglow spectra of 
GRB 190114C and GRB 180720B. This time again, using the same model, we successfully explain the multi-TeV afterglow spectra of GRB 190829A. Thus, our study shows that the three GRBs observed so far in the VHE $\gamma$-rays in the afterglow phases by IACTs can be explained very well in the context of the photohadronic model.

As is mentioned in the introduction, the VHE spectrum of GRB 180720B can be explained by the interactions of the Fermi accelerated protons with the synchrotron background, and the VHE spectra of GRB 190114C and GRB 190829A, by the interaction of protons with the SSC background. For the synchrotron background we should have $\delta < 2.0$ as the synchrotron photon flux is proportional to $\epsilon^{-\beta}$. On the other hand, for the SSC background, the photon flux is proportional to $\epsilon^{\beta}$, with $0 < \beta \le 1.0$ which corresponds to $2.5 \le \delta \le 3.0$. In future, through existing and forthcoming Cherenkov Telescopes, it might be possible to observe the VHE gamma-ray spectra from  nearby GRBs at redshift $\lesssim 0.5$ in their afterglow phases and some VHE gamma-ray spectra could only be explained by employing the two different zones, one with $\delta < 2.0$ and the another with $2.5 \le \delta \le 3.0$. Such a VHE spectrum can be interpreted as a result of the interactions of the high energy protons with both the synchrotron background and the SSC background in the forward shock region.

  We are thankful to Alberto Rosales de León, Gabriel Sánchez Colón and Benjamín Medina Carrillo for many useful discussions. The work of S.S. is partially supported by DGAPA-UNAM (Mexico) Projects No. IN103019 and No. IN103522. Partial support from CSU-Long Beach is gratefully acknowledged. 


\bibliography{grbref}{}
\bibliographystyle{aasjournal}

\end{document}